\documentclass[aps,pra,superscriptaddress,floatfix,showpacs,10pt,twocolumn]{revtex4}
\usepackage{amsmath}
\usepackage{graphicx}
\usepackage{amsfonts}
\usepackage{amssymb}

\usepackage[T1]{fontenc}
\usepackage[latin2]{inputenc}
\usepackage{latexsym}
\usepackage{bbm}
\usepackage{CJK}
\usepackage{color}

\begin{document}
\title{Geometric measure of quantum discord and the geometry of a class of two-qubit states}
\author{Wei Song}
\email{wsong1@mail.ustc.edu.cn(Corresponding~Author)}\affiliation{School
of Electronic and Information Engineering, Hefei Normal University,
Hefei 230061, China}
\author{Long-Bao Yu }
\affiliation{School of Electronic and Information Engineering, Hefei
Normal University, Hefei 230061, China}

\author{Ping Dong }
\affiliation{School of Electronic and Information Engineering, Hefei
Normal University, Hefei 230061, China}

\author{Da-Chuang Li}
\affiliation{School of Electronic and Information Engineering, Hefei
Normal University, Hefei 230061, China} \affiliation{Hefei National
Laboratory for Physical Sciences at the Microscale and Department of
Modern Physics, University of Science and Technology of China, Hefei
230026, China}

\author{Ming Yang}
\affiliation{School of Physics and Material Science, Anhui
University, Hefei 230039, China }
\author{Zhuo-Liang Cao}
\affiliation{School of Electronic and Information Engineering, Hefei
Normal University, Hefei 230061, China}

\date{\today}


\begin{abstract}
We investigate the geometric picture of the level surfaces of
quantum entanglement and geometric measure of quantum discord(GMQD)
of a class of X-states, respectively. This pictorial approach
provides us a direct understanding of the structure of entanglement
and GMQD. The dynamic evolution of GMQD under two typical kinds of
quantum decoherence channels is also investigated. It is shown that
there exists a class of initial states for which the GMQD is not
destroyed by decoherence in a finite time interval. Furthermore, we
establish a factorization law between the initial and final GMQD,
which allows us to infer the evolution of entanglement under the
influences of the environment.
\end{abstract}
\maketitle
\section{Introduction}
Entanglement is regarded as an ingredient resource for performing
almost all quantum information processing tasks\cite{Neilsen:2000}.
The situation started to change until a computational model was
presented named deterministic quantum computation with one
qubit(DQC1)\cite{Datta:2008}. Quantum discord was considered to be
the figure of merit for this model of quantum computation. Ever
since, quantum discord has attracted much
attention\cite{Ollivier:2001,Henderson:2001,Luo:2008,Modi:2010,Dakic:2010,Chen:2011,Shi:2011a,Shi:2011b,Cornelio:2011,Ferraro:2010,Streltsov:2011,Piani:2011,Al-Qasimi:2011,Bennett:2011,Galve:2011,Luo:2010,Hassan:2010,Yu:2011,Bylicka:2010,Huang:2011,Fanchini:2011,Zhang:2011a,Zhang:2011b,Shabani:2009,Bradler:2010}.
Quantum discord is a measure of nonclassical correlations between
two subsystems of a quantum system. It quantifies how much a system
can be disturbed when people observe it to gather information. Such
quantum correlations may be present in separable states and have a
non vanishing value for almost all quantum
states\cite{Ferraro:2010}. On the other hand, the quantum discord
can be used to indicate the quantum phase transitions better than
entanglement in certain physical systems at a finite temperature
\cite{Dillenschneider:2008,Werlang:2010}. In particular, two
operational interpretations of quantum discord have been proposed,
one in thermodynamics \cite{Zurek:2003} and the other from the
information theoretic perspective through the state merging protocol
\cite{Cavalcanti:2011,Madhok:2011}. These results establish the
status of quantum discord as another important resource for quantum
informational processing tasks besides entanglement.

Originally, the first definition of quantum discord was given by
Ollivier and Zurek \cite{Ollivier:2001}and, independently, by
Henderson and Vedral \cite{Henderson:2001}. The quantum discord of a
composite system AB is defined by $D_A  \equiv \mathop {\min
}\limits_{\left\{ {E_i^A } \right\}} \sum\limits_i {p_i } S\left(
{\rho _{B|i} } \right) + S\left( {\rho _A } \right) - S\left( {\rho
_{AB} } \right)$, where $S\left( {\rho _{AB} } \right) = Tr\left(
{\rho _{AB} \log _2 \rho _{AB} } \right)$ is the von Neumann entropy
and the minimum is taken over all positive operator valued
measures(POVMs) $\left\{ {E_i^A } \right\}$ on the subsystem $A$
with $p_i  = Tr\left( {E_i^A \rho _{AB} } \right)$ being the
probability of the i-th outcome and $\rho _{B|i}  = {{Tr_A \left(
{E_i^A \rho _{AB} } \right)} \mathord{\left/
 {\vphantom {{Tr_A \left( {E_i^A \rho _{AB} } \right)} {p_i }}} \right.
 \kern-\nulldelimiterspace} {p_i }}$ being the conditional state of
 subsystem $B$. In a more restrict sense, the minimum is often taken over the von Neumann
 measurements. However, it is notoriously difficult to compute because
of the minimization taken over all possible POVM, or von Neumann
measurements. At present, there are only a few analytical results
including the Bell-diagonal states \cite{Luo:2008}, rank-2 states
\cite{Shi:2011b,Cen:2011}and Gaussian states
\cite{Giorda:2010,Adesso:2010}. In addition, a simple algorithm to
evaluate the quantum discord for two-qubit X-states is proposed by
Ali \emph{et al.} \cite{Ali:2010} with minimization taken over only
a few cases. Unfortunately, their algorithm is valid only for a
family of X-states\cite{Chen:2011,Lu:2011a}. Recently Shi \emph{et
al.} \cite{Shi:2011c} present an efficient method to solve this
problem. For the general two-qubit states, the evaluation of quantum
discord remains a nontrivial task and only some lower and upper
bounds are available \cite{Yu:2011c}. In order to avoid the
difficulties in minimization procedures a geometric view of quantum
discord was introduced. Generally, there are two versions of
geometric measure of quantum discord(GMQD). In the first version the
concept of relative entropy is used as a distance measure of
correlations \cite{Modi:2010}. The second version is defined by the
Hilbert-Schmidt norm measure \cite{Dakic:2010}. The
relative-entropy-based discords have the drawback that their
analytical expressions are known only for certain limited classes of
states. Below we only consider the second version of GMQD.
Especially, Dakic \emph{et al} \cite{Dakic:2012} show that the GMQD
is related to the fidelity of remote state preparation which
provides an operational meaning to GMQD. Formerly, this geometric
measure of quantum discord is defined by $ D_A^g  = \mathop {\min
}\limits_{\chi  \in \Omega _0 } \left\| {\rho - \chi } \right\|^2 $,
where $ \Omega _0 $ denotes the set of zero-discord states and $
\left\| {X - Y} \right\|^2  = Tr\left( {X - Y} \right)^2 $ is the
square norm in the Hilbert-Schmidt space. The subscript $A$ denotes
that the measurement is taken on the system $A$. An arbitrary
two-qubit state can be written in Bloch representation:

\begin{eqnarray}
 \rho  = \frac{1} {4}\left[ {I \otimes I + \sum\limits_i^3 {\left(
{x_i \sigma _i  \otimes I + y_i I \otimes \sigma _i } \right) +
\sum\limits_{i,j = 1}^3 {R_{ij} \sigma _i  \otimes \sigma _j } } }
\right]
\end{eqnarray}

\noindent where $ x_i  = Tr\rho \left( {\sigma _i  \otimes I}
\right),y_i  = Tr\rho \left( {I \otimes \sigma _i } \right) $ are
components of the local Bloch vectors, $ \sigma _i ,i \in \left\{
{1,2,3} \right\} $ are the three Pauli matrices, and $ R_{ij}  =
Tr\rho \left( {\sigma _i  \otimes \sigma _j } \right) $ are
components of the correlation tensor. For the two-qubit case, the
zero-discord state is of the form $ \chi  = p_1 \left| {\psi _1 }
\right\rangle \left\langle {\psi _1 } \right| \otimes \rho _1  + p_2
\left| {\psi _2 } \right\rangle \left\langle {\psi _2 } \right|
\otimes \rho _2$, where $ \left\{ {\left| {\psi _1 } \right\rangle
,\left| {\psi _2 } \right\rangle } \right\} $ is a single-qubit
orthonormal basis. Then an analytic expression of the GMQD is given
by \cite{Dakic:2010}:

\begin{eqnarray}
D_A^g \left( \rho  \right) = \frac{1} {4}\left( {\left\| x
\right\|^2  + \left\| R \right\|^2  - k_{\max } } \right)
\end{eqnarray}

\noindent where $ x = \left( {x_1 ,x_2 ,x_3 } \right)^T $ and $
k_{\max } $ is the largest eigenvalue of matrix $ K = xx^T  + RR^T
$.

\begin{figure}[ptb]
\includegraphics[scale=0.70,angle=0]{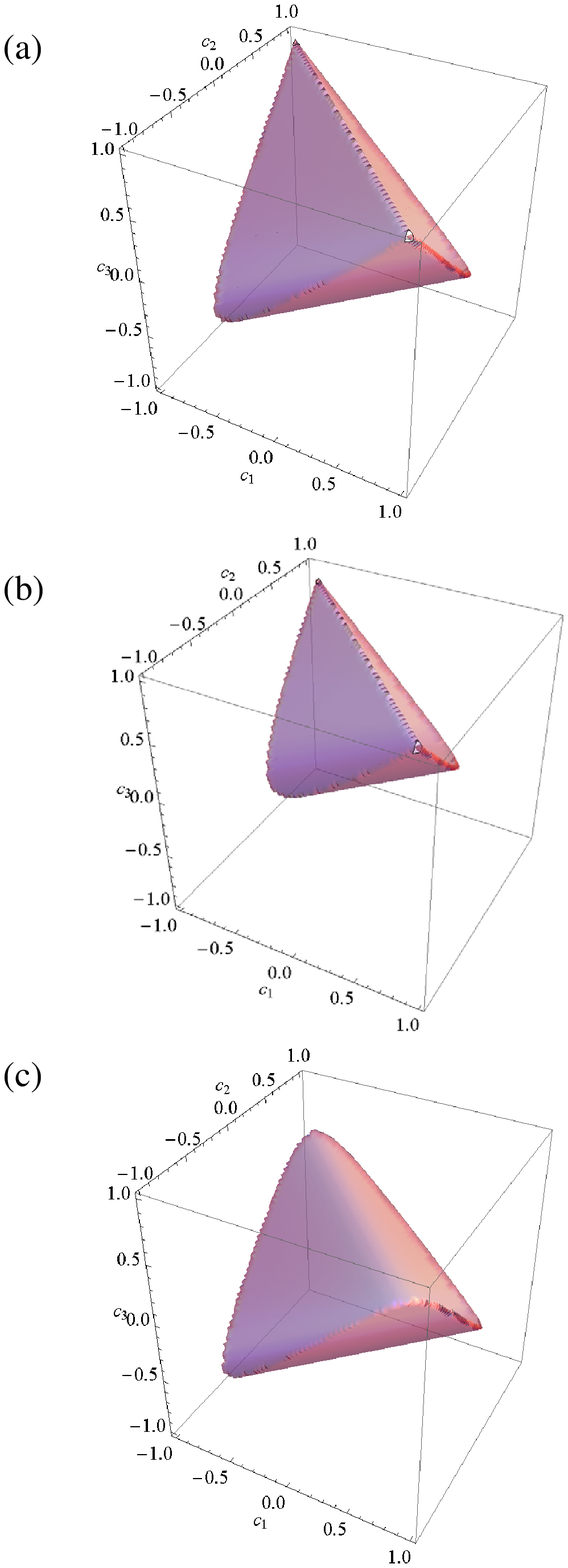}\caption{(Color online). The geometry of the set of valid
states with different $r$ and $s$, repectively.
(a)$r$=$s$=0.3,(b)$r$=$s$=0.5,(c)$r$=0.4,$s$=0.1}
\label{fig1}%
\end{figure}

In this paper we shall investigate the level surfaces of
entanglement and GMQD for a class of two-qubit states using the
geometric picture presented in \cite{Lang:2010}. It is well known
that the set of Bell-diagonal states for two qubits can be depicted
as a tetrahedron in three dimensions in Bloch representation
\cite{Horodecki:2009}. Analogous to entanglement, Lang and Caves
have depicted the level surfaces of quantum discord for
Bell-diagonal states. More recently, Girolami and Adesso
\cite{Girolami:2011} and independently Batle \emph{et al.}
\cite{Batle:2011} provided numerical evidence, from which one can
infer that discord and GMQD may be different. It is thus worth
investigating the GMQD from the geometric picture. In this sense,
our research provides a direct understanding of the structure of
GMQD. We consider a class of X-states that the Bloch vectors are $z$
directional, which include Bell-diagonal states as a special case.
We show that the level surface of GMQD is very different from the
quantum discord. On the other hand, the dynamics of quantum discord
has attracted much attention due to the inevitable interaction with
environment. The dynamic behavior of GMQD is also investigated under
two typical kinds of decoherence channels. We find a class of states
for which the GMQD is not destroyed by decoherence in a finite time
interval. Interestingly, we also obtain a factorization law for GMQD
which allows us to infer the evolution of entanglement under the
influences of the environment, \emph{e.g.} the phase damping
channel(PDC), and the depolarizing channel(DPC).

The paper is organized as follows. In Sec.II, we present the
geometric picture of the level surfaces of quantum entanglement and
GMQD of a class of X-states with $z$ directional Bloch vectors,
respectively. It is shown that the surface of constant GMQD varies
with the local Bloch vectors. In Sec.III, we investigate the dynamic
evolution of GMQD under quantum decoherence channels, and obtain
analytic results for two typical kinds of quantum decoherence
channels. A summary is given in Sec.IV.

\section{geometrical picture of entanglement and GMQD}

For analytical simplicity, we consider the following two-qubit X
states:

\begin{eqnarray}
\rho  = \frac{1} {4}\left[ {I \otimes I + \textbf{r} \cdot \sigma
\otimes I + I \otimes \textbf{s} \cdot \sigma  + \sum\limits_{i =
1}^3 {c_i \sigma _i \otimes \sigma _i } } \right]
\end{eqnarray}

\noindent where we choose the Bloch vectors as $z$ directional with
$ \textbf{r} = \left( {0,0,r} \right)$,$ \textbf{s} = \left( {0,0,s}
\right) $. The GMQD can be calculated explicitly for this state,
thus allowing us to get analytic results. If
$\textbf{r}=\textbf{s}=0$, $ \rho $ is reduced to the two-qubit
Bell-diagonal states. Horodecki have shown that Bell-diagonal states
belongs to a tetrahedron with vertices
$(1,-1,1),(-1,1,1),(1,1,-1)$,and $ (-1,-1,-1)$ in the Bloch
representation. From the positivity of the eigenvalues of $ \rho $
in Eq.(3), we have

\begin{eqnarray}
0 \leqslant \frac{{\text{1}}} {{\text{4}}}\left( {1 - \sqrt {r^2  -
2rs + s^2  + c_1^2  + 2c_1 c_2  + c_2^2 }  - c_3 } \right) \leqslant
1,\nonumber\\
0 \leqslant \frac{{\text{1}}} {{\text{4}}}\left( {1 + \sqrt {r^2  -
2rs + s^2  + c_1^2  + 2c_1 c_2  + c_2^2 }  - c_3 } \right) \leqslant
1,\nonumber\\
0 \leqslant \frac{{\text{1}}} {{\text{4}}}\left( {1 - \sqrt {r^2  +
2rs + s^2  + c_1^2  - 2c_1 c_2  + c_2^2 }  + c_3 } \right) \leqslant
1,\nonumber\\
0 \leqslant \frac{{\text{1}}} {{\text{4}}}\left( {1 + \sqrt {r^2  +
2rs + s^2  + c_1^2  - 2c_1 c_2  + c_2^2 }  + c_3 } \right) \leqslant
1\nonumber\\
\end{eqnarray}

\noindent For fixed parameters $r$ and $s$, the above inequalities
become a three-parameter set, whose geometry can be depicted in the
three dimensional correlation state space. In Fig.1 we plot the
physical region with different $r$ and $s$, respectively. Fig.1
shows that physical regions of the state $ \rho $ shrink with larger
$r$ and $s$. We plot in Fig.2 the level surfaces of constant
concurrence with fixed $r$ and $s$ for three cases. Here, we choose
concurrence to measure entanglement which is defined as $ C = \max
\left\{ {0,\lambda _1  - \lambda _2  - \lambda _3  - \lambda _4 }
\right\}$, where the $ \lambda _i $ are, in decreasing order, the
square roots of the eigenvalues of the matrix $ \rho \sigma _y
\otimes \sigma _y \rho ^* \sigma _y \otimes \sigma _y $ where $ \rho
^* $ is the complex conjugate of $ \rho $. As shown in Fig.2, the
level surfaces of constant concurrence for the state $\rho $ defined
in Eq.(3) consist of four discrete pieces, and the areas decrease
when the concurrence increases.An extremal case is the four vertices
$(1,-1,1),(-1,1,1),(1,1,-1)$, and $ (-1,-1,-1)$ of the tetrahedron
corresponding to the four Bell states with maximal concurrence.
Finally, we investigate the GMQD from the geometric picture. For the
state $ \rho $, the GMQD can be calculated in the method presented
in Ref. \cite{Lu:2010c} By introducing a matric $\mathcal{R}$
defined by

\begin{equation}
\label{eq5} \mathcal{R}  = \left( {{\begin{array}{*{20}c}
1 & {y^T }  \hfill \\
x & R \hfill \\
\end{array} }} \right)
\end{equation}
\noindent and $ 3 \times 4$ matric $\mathcal{ {R'} }$ through
deleting the first row of $\mathcal{R}$, then the GMQD is given by

\begin{eqnarray}
D_A^g \left( \rho  \right) = \frac{1} {4}\left[ {\left(
{\sum\limits_k {\lambda _{_k }^2 } } \right) - \mathop {\max
}\limits_k \lambda _{_k }^2 } \right]
\end{eqnarray}
\noindent where $ {\lambda _k } $ is the singular values of
$\mathcal{R'}$. For the two-qubit state $\rho $ in Eq.(3), we obtain
\begin{eqnarray}
D_A^g \left( \rho  \right)=\frac{1} {4}\left( {c_1^2  + c_2^2  +
c_3^2  + r^2  -  {\text{Max}} \left( {c_1^2 ,c_2^2 ,c_3^2  + r^2 }
\right)} \right)
\end{eqnarray}

The geometric picture is depicted in terms of the constant GMQD in
Fig.3. From these plots one can see that the shape of the constant
GMQD is quite different from quantum discord. It also shows
different shapes for different local Bloch vectors. The constant
surfaces are cut off by the physical region of state $ \rho $. For
small discord the surface is continuous, and it becomes discrete
pieces for larger discord. At the four vertices
$(1,-1,1),(-1,1,1),(1,1,-1)$,and $ (-1,-1,-1)$ of the tetrahedron
the GMQD reaches its maximal value. Furthermore, one can see that
GMQD is neither concave nor convex as shown in Fig.3.

\begin{figure}[ptb]
\includegraphics[scale=0.70,angle=0]{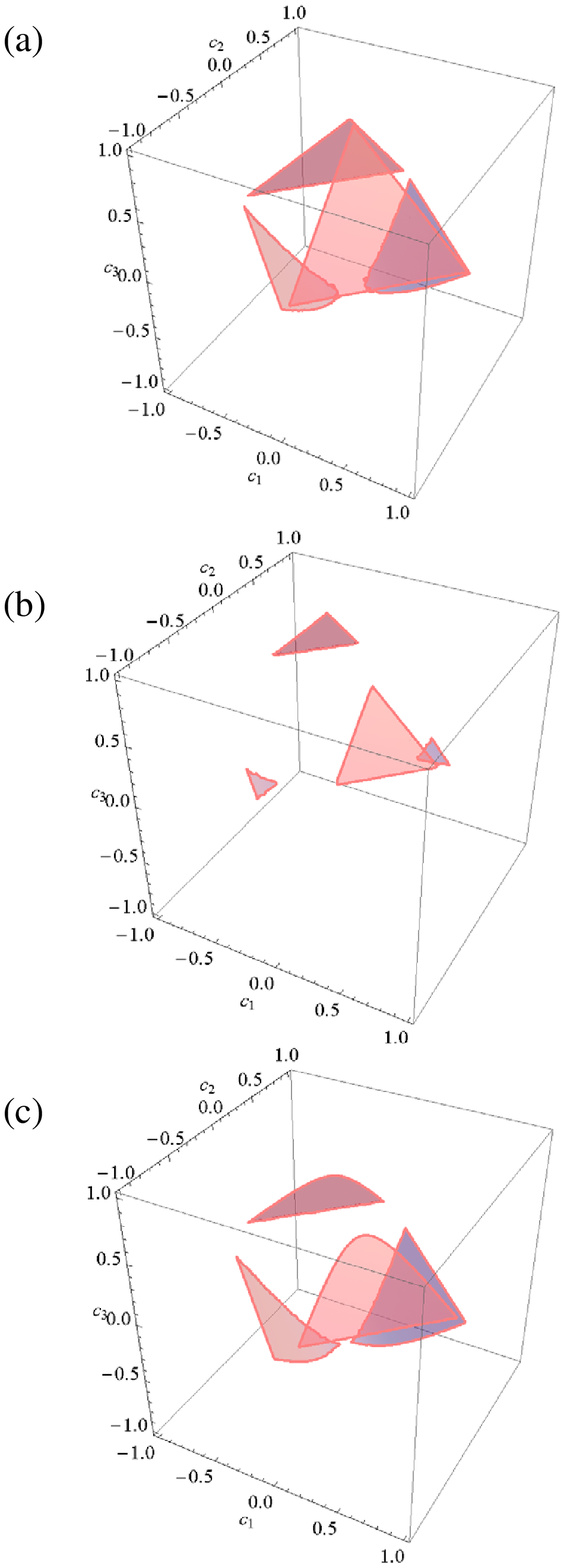}\caption{(Color online). Surfaces of constant concurrence.
(a)$r$=$s$
=0.3,$C(\rho)$=0.03,(b)$r$=$s$=0.5,$C(\rho)$=0.35,(c)$r$=0.4,$s$=0.1,$C(\rho)$=0.03}
\label{fig1}%
\end{figure}

\begin{figure}[ptb]
\includegraphics[scale=0.70,angle=0]{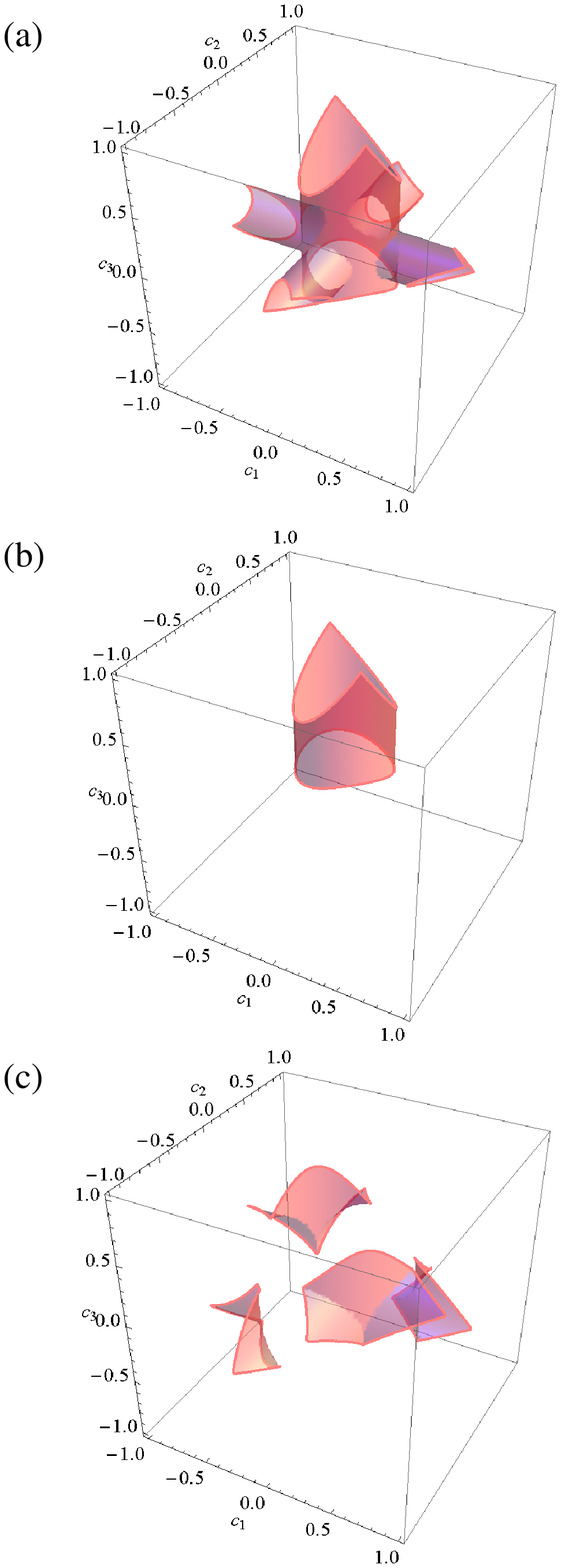}\caption{(Color online). Surfaces of constant GMQD.
(a)$r$=$s$
=0.3,$D(\rho)$=0.03,(b)$r$=$s$=0.5,$D(\rho)$=0.35,(c)$r$=0.4,$s$=0.1,$D(\rho)$=0.08}
\label{fig1}%
\end{figure}

\section{Dynamics of GMQD under local decoherence channels}

In this section we consider the state affected by the action of two
independent channels and calculate the GMQD analytically. The
dynamics of quantum discord has been investigated in both Markovian
and non-Markovian environments and has been demonstrated
experimentally
\cite{Werlang:2009b,Maziero:2009,Wang:2010,Mazzola:2010,Li:2011,Fanchini:2010,Auccaise:2011,Xu:2010a,Xu:2010b}.
It has been shown that the behaviors of quantum discord and GMQD may
be different. It is thus desirable to consider the evolution of GMQD
under different decoherence channels. Here, we consider two typical
kinds of decoherence channels: the phase damping channel(PDC), and
the depolarizing channel(DPC). To calculate the dynamics of GMQD, we
turn to the Heisenberg picture to describe the quantum channels. In
order to obtain the analytic expressions of GMQD of the state
subject to local decoherence channels, we need to calculate the
expection matrix $\mathcal{R}$. In the Heisenberg picture
\cite{Lu:2010c,Wang:2010d}, the expectation matrix $\mathcal{R}$ is
given by

\begin{eqnarray}
\mathcal{R}_{ij}  = \left( {M_A \mathcal{R}_0 M_B^T } \right)_{ij}
\end{eqnarray}

\noindent where $ \mathcal{R}_0  = Tr\left( {\sigma _i  \otimes
\sigma _j \rho _0 } \right) $, $i \in \left\{ {0,1,2,3} \right\} $,
$ \rho _{\text{0}} $ is the initial state, and $ M_{A\left( B
\right)} $ is the transmission matrix of each local channel. For
simplicity, we choose the local channels to be identical. In this
case,  the transmission matrices can be written as

\begin{equation}
\label{eq5} M_{PDC}  = \left( {{\begin{array}{*{20}c}
1 & 0 & 0  & 0 \hfill \\
0 & 1-p & 0  & 0 \hfill \\
0 & 0 & 1-p  & 0 \hfill \\
0 & 0 & 0  & 1 \hfill \\

\end{array} }} \right),
\end{equation}

\begin{equation}
\label{eq5} M_{DPC}  = \left( {{\begin{array}{*{20}c}
1 & 0 & 0  & 0 \hfill \\
0 & 1-p & 0  & 0 \hfill \\
0 & 0 & 1-p  & 0 \hfill \\
0 & 0 & 0  & 1-p \hfill \\

\end{array} }} \right),
\end{equation}

\noindent For state (3), $ \mathcal{R}_0  = \left\{ {\left\{
{1,0,0,s} \right\},\left\{ {0,c_1 ,0,0} \right\},\left\{ {0,0,c_2
,0} \right\}} \right. $\\$ \left. {\left\{ {r,0,0,c_3 } \right\}}
\right\}$. According to the above formula, we have

\begin{eqnarray}
D_{PDC}^g  = \frac{1} {4}\left[ {\left( {1 - p} \right)^4 c_1^2  +
\left( {1 - p} \right)^4 c_2^2  + r^2  + c_3^2 } \right.\nonumber\\
\left. { - \max \left\{ {\left( {1 - p} \right)^4 c_1^2 ,\left( {1 -
p} \right)^4 c_2^2 ,r^2  + c_3^2 } \right\}} \right]
\end{eqnarray}

\begin{eqnarray}
D_{DPC}^g= \frac{1} {4}\left[ {\left( {1 - p} \right)} \right.^4
c_1^2  + \left( {1 - p} \right)^4 c_2^2\nonumber\\
+ \left( {1 - p} \right)^2 r^2  + \left( {1 - p} \right)^4 c_3^2
\nonumber\\
- \max \left\{ {\left( {1 - p} \right)^4 } \right.c_1^2 ,\left( {1 -
p} \right)^4 c_2^2\nonumber\\
\left. {\left. {\left( {1 - p} \right)^2 r^2  + \left( {1 - p}
\right)^4 c_3^2 } \right\}} \right]
\end{eqnarray}

For some Bell-diagonal states, it has been shown that quantum
discord is not destroyed by decoherence for some finite time
interval \cite{Mazzola:2010}. A natural question arises: Whether
such a phenomena exists for GMQD? We consider the state in Eq.(3)
undergoes two identical PDCs. In this case, $p = 1 - \exp ( - \gamma
t)$, where $\gamma $ is the phase damping rate. For $c_1 = 0,c_2^2
> r^2  + c_3^2 $, suppose ${p_1 }$ satisfies the equation $\left( {1 - p_1 } \right)^4 c_2^2  = r^2  + c_3^2
$. If ${p < p_1 }$, from Eq.(11) we have $D_{PDC}^g  = \frac{1}
{4}\left( {r^2  + c_3^2 } \right)$ which is independent of time.
Therefore, we conclude that for a finite time interval the GMQD does
not decay despite the presence of local phase damping noises. It is
directly seen that such a phenomena also exists for the case $c_2 =
0,c_1^2
> r^2  + c_3^2 $. These results show that GMQD remains intact under the action of some special kinds of quantum channels.
It should be noted that similar phenomenon have also been noticed by
Karpat \emph{et al} \cite{Karpat:2011} for the qubit-qutrit systems.
In the geometric picture, this behavior corresponds to the state
evolving along a straight line in the constant GMQD tube until it
enconters another constant GMQD tube.

Hitherto, we have only considered the time evolution of GMQD under
PDC or DPC described by Eq.(11) and Eq.(12). Next, we want to derive
a more general result on GMQD relating the initial and final state
of GMQD. Inspired by the famous factorization law for entanglement
decay derived by Konrad \emph{et al.} \cite{Konrad:2008}, we find an
analogous factorization law between the initial and final GMQD of
the class of two-qubit states defined in Eq.(3) subject to two
different local decoherence channels. We state our result as the
following theorem.

\noindent \textbf{Theorem. }Consider the class of X-states defined
in Eq.(3), with each qubit being subject to the local decoherence
channels, \emph{i.e.} the phase damping channel(PDC) or the
depolarizing channel(DPC). The time evolution of GMQD satisfies
\begin{eqnarray}
D^g \left[ {\left( {\$ _1  \otimes \$ _2 } \right)\rho(t) } \right]
\geqslant 2D^g \left[ {\left( {\$ _1  \otimes \$ _2 } \right)\left|
{\beta _i } \right\rangle \left\langle {\beta _i } \right|}
\right]D^g \left[ {\rho }(0) \right].
\end{eqnarray}

\noindent where the local decoherence channels are represented by $
{\$ _1 } $ and ${\$ _2 } $, $ \rho _0 $ is the initial state and $
{\left| {\beta _i } \right\rangle }, i \in \left\{ {1,2,3,4}
\right\} $ denotes one of the four Bell states.

\noindent \textbf{Proof.} First we consider the state $\rho(0) $ is
effected by the action of two identical local PDCs,  the time
evolution of GMQD is given by Eq.(11). For convenience we divide the
proof into three cases.

\noindent Case 1. $ r^2  + c_3^2  \geqslant \left\{ {c_1^2 ,c_2^2 }
\right\} $. In this case, it is easy to show that the inequality
becomes equality.

\noindent Case 2. $ c_1^2  \geqslant c_2^2  \geqslant r^2  + c_3^2$.
First, using Eq.(7) we have $ D^g \left( {\rho(0) } \right) =
\frac{1} {4}\left( {c_2^2  + r^2  + c_3^2 } \right) $. Suppose $
{p_0 } $ satisfies the equation $ \left( {1 - p_0 } \right)^4 c_1^2
= r^2  + c_3^2 $. If $ p \leqslant p_0 $, the GMQD of the state
$\rho $ is given by

\begin{eqnarray}
D_{^{PDC} }^g \left( \rho (0) \right) &=& \frac{1} {4}\left[ {\left(
{1 -
p} \right)^4 c_2^2  + r^2  + c_3^2 } \right]\nonumber\\
 &\geqslant& \frac{1}
{4}\left[ {\left( {1 - p} \right)^4 c_2^2  + \left( {1 - p}
\right)^4 \left( {r^2  + c_3^2 } \right)} \right]\nonumber\\
 &= &\left( {1 - p} \right)^4 D^g \left( {\rho(0)  } \right)\nonumber\\
 &=& 2D^g \left[ {\left( {\$ _{PDC}  \otimes \$ _{PDC} } \right)\left|
{\beta _i } \right\rangle \left\langle {\beta _i } \right|}
\right]D^g \left( {\rho(0) } \right)\nonumber\\
\end{eqnarray}

If $ p > p_0$, then
\begin{eqnarray}
D_{^{PDC} }^g \left( \rho (t) \right) &=& \frac{1} {4}\left[ {\left(
{1 -
p} \right)^4 c_1^2  + \left( {1 - p} \right)^4 c_2^2 } \right]\nonumber\\
&\geqslant& \frac{1} {4}\left[ {\left( {1 - p} \right)^4 c_2^2  +
\left( {1 - p} \right)^4 \left( {r^2  + c_3^2 } \right)} \right]
\nonumber\\
&=& \left( {1 - p} \right)^4 D^g \left( {\rho(0) } \right)\nonumber\\
&=& 2D^g \left[ {\left( {\$ _{PDC}  \otimes \$ _{PDC} }
\right)\left| {\beta _i } \right\rangle \left\langle {\beta _i }
\right|}
\right]D^g \left( {\rho(0) } \right)\nonumber\\
\end{eqnarray}

\noindent Case 3. $ c_1^2  \geqslant r^2  + c_3^2  \geqslant c_2^2
$. For $ p \leqslant p_0 $ or $ p
> p_0$, the proof is similar to case 2.

By simply exchanging $ c_1 $ and $ c_2$ we can verify the above
relations for the  cases $ c_2^2  \geqslant c_1^2 \geqslant r^2  +
c_3^2 $ and $ c_2^2  \geqslant r^2  + c_3^2 \geqslant c_1^2 $. For
the case of two identical local DPCs, one can prove the above
results in the same way as for PDCs. In the following, we consider
the first qubit is subject to the PDC and the second qubit is
subject to the DPC. The expectation matrix $\mathcal{R}$ can be
calculated according to the formula $ \mathcal{R} = M_A
\mathcal{R}_0 M_B^T $, where $ M_{A\left( B \right)} $ is the
transformation matrix of PDC(DPC). Thus, the GMQD is given by
\begin{eqnarray}
D^g \left( {\rho \left( t \right)} \right) = \frac{1} {4}\left[
{\left( {1 - p} \right)} \right.^4 c_1^2  + \left( {1 - p} \right)^4
c_2^2+ r^2  + \left( {1 - p} \right)^2 c_3^2\nonumber\\
\left. {{-\text{ max}}\left\{ {\left( {1 - p} \right)^4 c_1^2
,\left( {1 - p} \right)^4 c_2^2 ,r^2  + \left( {1 - p} \right)^2
c_3^2 } \right\}} \right]
\nonumber\\
\end{eqnarray}
\noindent where we have assumed that the parameter $p$ is the same
in the two decoherence channels. Then it suffices to consider three
separate cases.

\noindent Case 1. $r^2  + c_3^2 \geqslant \left\{ {c_1^2 ,c_2^2 }
\right\} $. Then
\begin{eqnarray} D^g \left( {\rho \left( t
\right)} \right) = \frac{1} {4}\left[ {\left( {1 - p} \right)^4
c_1^2  + \left( {1 - p} \right)^4 c_2^2 } \right]\nonumber\\
= 2D^g \left[ ({\$ _{PDC} \otimes \$ _{DPC} )\left( {\left| {\beta
_i } \right\rangle \left\langle {\beta _i } \right|} \right)}
\right]D^g \left( {\rho \left( 0 \right)} \right)
\end{eqnarray}

\noindent Case 2. $ {c_1^2  \geqslant c_2^2  \geqslant r^2  + c_3^2
} $. Suppose $ p'_0 $ satisfies $ \left( {1 - p'_0} \right)^4 c_1^2
= r^2 + \left( {1 - p'_0} \right)^2 c_3^2 $. If $ p \leqslant p'_0
$, we have

\begin{eqnarray}
D^g \left( {\rho \left( t \right)} \right) = \frac{1} {4}\left[
{\left( {1 - p} \right)^4 c_2^2  + r^2  + \left( {1 - p} \right)^2
c_3^2 } \right]\nonumber\\
 \geqslant \frac{1}
{4}\left[ {\left( {1 - p} \right)^4 c_2^2  + \left( {1 - p}
\right)^4 r^2  + \left( {1 - p} \right)^4 c_3^2 } \right]
\nonumber\\
 = 2D^g \left[ ({\$ _{PDC}  \otimes \$ _{DPC}) \left( {\left| {\beta _i } \right\rangle \left\langle {\beta _i } \right|} \right)} \right]D^g \left( {\rho \left( 0 \right)} \right)
\end{eqnarray}

If $ p > p'_0 $, we obtain

\begin{eqnarray}
D^g \left( {\rho \left( t \right)} \right) = \frac{1} {4}\left[
{\left( {1 - p} \right)^4 c_1^2  + \left( {1 - p} \right)^4 c_2^2 }
\right]
\nonumber\\
 \geqslant \frac{1}
{4}\left( {1 - p} \right)^4 \left[ {c_2^2  + r^2  + c_3^2 } \right]
\nonumber\\
 = 2D^g \left[ ({\$ _{PDC}  \otimes \$ _{DPC}) \left( {\left| {\beta _i } \right\rangle \left\langle {\beta _i } \right|} \right)} \right]D^g \left( {\rho \left( 0 \right)} \right)
\end{eqnarray}

\noindent Case 3. $ {c_1^2  \geqslant r^2  + c_3^2  \geqslant c_2^2
} $. In this case, the proof is similar to case 2.

One can also directly verify that the above relation holds for the
other cases.$\hfill\blacksquare$

The theorem above provides us a method to compute the lower bound of
the time evolution of a class of X-states under two typical kinds of
decoherence channels, without resorting to the time evolution of the
underlying quantum state itself. This inequality also holds for the
one-sided PDC or DPC, which is summarized as the following
Corollary:

\noindent \textbf{ Corollary 1}. For the class of X-states defined
in Eq.(3), with one qubit being subject to PDC or DPC, we have

\begin{eqnarray}
D^g \left( {\rho \left( t \right)} \right) \geqslant 2D^g \left[
({\$ _i  \otimes I)\left( {\left| {\beta _i } \right\rangle
\left\langle {\beta _i } \right|} \right)} \right]D^g \left( {\rho
\left( 0 \right)} \right)
\end{eqnarray}

Moreover, for $r=s=0$, \emph{i.e.} Bell diagonal states, using
Eq.(12), one can directly calculate the above inequality becomes
equality which we summarize as follows:

\noindent \textbf{ Corollary 2}. For arbitrary Bell-diagonal states
subject to two DPCs, the evolution of GMQD is given by

\begin{eqnarray}
D_{DPC}^g \left( \rho (t) \right) =2D^g \left[ {\left( {\$ _{DPC}
\otimes \$ _{DPC} } \right)\left| {\beta _i } \right\rangle
\left\langle {\beta _i } \right|} \right]D^g \left( \rho (0) \right)\nonumber\\
\end{eqnarray}

\noindent where the two DPCs may be different. By far we have
considered the time evolution of GMQD in the presence of two typical
kinds of local decoherence noise. Another important decoherence
noise is the amplitude damping channel(ADC). Evidence can show that
the above relation also holds for two identical ADCs, i.e.

\begin{eqnarray}
D^g \left( {\rho \left( t \right)} \right) \geqslant 2D^g \left[
({\$ _{ADC}  \otimes \$ _{ADC} )\left( {\left| {\beta _i }
\right\rangle \left\langle {\beta _i } \right|} \right)} \right]D^g
\left( {\rho \left( 0 \right)} \right)\nonumber\\
\end{eqnarray}

In Fig.4, we plot the evolution of GMQD under two identical ADCs and
its lower bound in Eq.(22) when (a)$ c_1 = 0.1,c_2 = 0.1,c_3 = 0.2,r
= s = 0.3 $; (b)$ c_1  = 0.2,c_2  = 0.05,c_3  = 0.3,r = 0.4,s =
0.1$.
\begin{figure}[ptb]
\includegraphics[scale=0.45,angle=0]{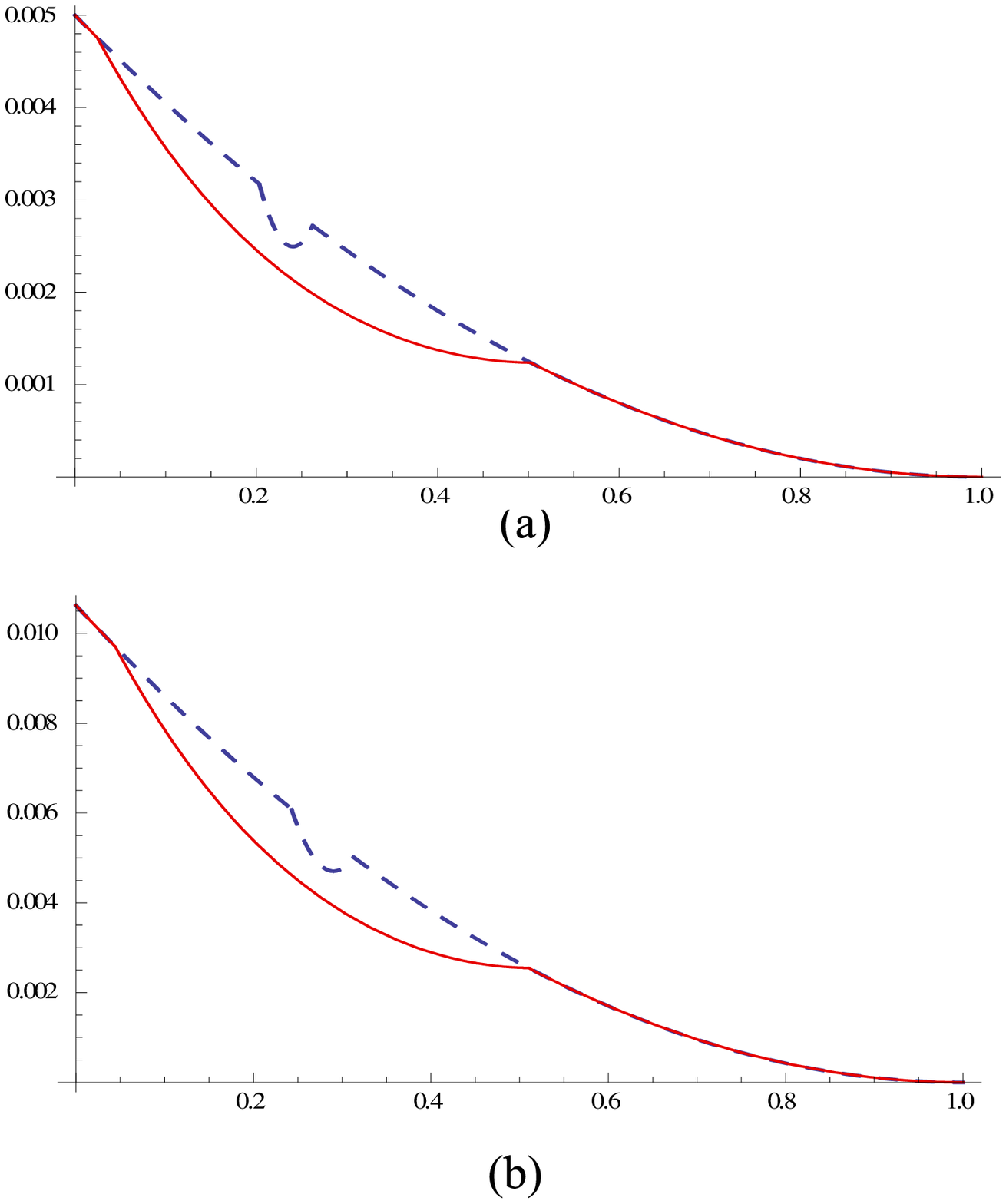}\caption{(Color online).Plots of the dynamics of GMQD under two
identical ADCs(blue line) and its lower bound defined in Eq.(22)(red
line).(a)$c_1 = 0.1,c_2  = 0.1,c_3 = 0.2,r = s = 0.3 $; (b)$ c_1  =
0.2,c_2  = 0.05,c_3  = 0.3,r = 0.4,s = 0.1$.}
\label{fig1}%
\end{figure}

\section{DISCUSSIONS AND CONCLUSIONS}

In this work, we investigated the level surfaces of GMQD for a class
of two-qubit X-states from the geometric picture. First, we plot the
physical region for a class of two-qubit X-states with fixed local
Bloch vectors. It is shown that physical regions of the state have
different geometry with the Bell-diagonal states and shrink with
larger Bloch vectors. Second, the geometric picture is depicted in
terms of the constant concurrence and GMQD, respectively. We find
that the shape of the surfaces has close relationship with the value
of GMQD and local Bloch vectors. Finally, we also investigate the
dynamics of GMQD under two typical kinds of decoherence channels and
obtain analytic results of the evolution of GMQD. It is shown that
there exists a class of initial states for which the GMQD is not
destroyed by decoherence in a finite time interval. Moreover, a
direct factorization relationship between the initial and final GMQD
subject to two typical kinds of decoherence channels is derived.
This factorization law allows us to infer the evolution of
entanglement under the influences of the environment without
resorting to the time evolution of the initial quantum state itself.
An open question is whether this law holds under general local
decoherence channels. Our results imply that further study on the
dynamics of GMQD is required.

\section{ACKNOWLEDGMENTS}

We are grateful to the referee for valuable suggestions. This work
was supported by the National Natural Science Foundation of China
under Grant No.10905024,No.11005029,No.11104057,No.11204061, the Key
Project of Chinese Ministry of Education under Grant No.211080, and
the Key Program of the Education Department of Anhui Province under
Grant No.KJ2011A243, No.KJ2012A244, No.KJ2012A245, the Anhui
Provincial Natural Science Foundation under Grant No.11040606M16,
No.10040606Q51, the Doctoral Startup Foundation of Hefei Normal
University under Grant No.2011rcjj03.

Note added. After completing this manuscript, we became aware of an
interesting related works by Yao Yao \emph{et al}\cite{Yao:2011}
recently.

\end{document}